\documentclass[a4paper,twocolumn,
english,aps,prb,floatfix,showpacs]{revtex4}
\usepackage[T1]{fontenc}
\usepackage[latin1]{inputenc}
\usepackage{amsmath}
\usepackage{babel}
\usepackage{graphics}
\usepackage{amssymb}


\begin{document}
 
\title{Failure of single-parameter scaling of wave functions in \\
Anderson localization}

\author{S.L.A. \surname{de Queiroz}}

\email{sldq@if.ufrj.br}

\affiliation{Instituto de F\'\i sica, Universidade Federal do
Rio de Janeiro, Caixa Postal 68528, 21945-970
Rio de Janeiro RJ, Brazil}

\date{\today}

\begin{abstract}
We show how to use properties of the vectors which are iterated 
in the transfer-matrix approach to Anderson 
localization,  in order to generate the statistical distribution of 
electronic  wavefunction amplitudes at arbitrary distances from the origin 
of $L^{d-1} \times \infty$ disordered systems. For $d=1$ our approach 
is shown to reproduce exact diagonalization results available
in the literature. 
In $d=2$, where strips of width $ L \leq 64$ sites were  used, attempted
fits of gaussian (log-normal) forms to the wavefunction amplitude 
distributions result 
in effective localization lengths growing with distance, contrary to
the prediction from single-parameter scaling theory. We also 
show that the distributions possess a negative skewness $S$, which  
is invariant under the usual histogram-collapse rescaling, and
whose absolute value increases with distance. We find $0.15 \lesssim -S 
\lesssim 0.30$ for the range of parameters used in our study.
\end{abstract}
\pacs{PACS numbers:  71.23.An, 73.20.Fz}
\maketitle
\section{INTRODUCTION}
\label{intro}
The localization model introduced by Anderson~\cite{and58}
incorporates two basic elements, namely the rules of quantum mechanics
applied to a single-electron, tight-binding model Hamiltonian, plus
quenched disorder (realized, e.g., by assigning random self-energies to
lattice sites). Its original purpose was to
show the existence of a  disorder-induced transition in 
three-dimensional systems, from metallic (diffusive)
to  insulating (localized) electronic behavior, upon increasing
randomness. Over the years the model has turned out to exhibit a rich
variety of physical aspects, many of them highlighted by 
the (single--parameter) scaling theory of 
localization (SPST)~\cite{gof4,mkk93}. 
Of particular interest here is the  fact
that, in zero magnetic field and in the absence of spin-orbit couplings, 
SPST predicts insulating behavior, for any finite amount of 
disorder, in spatial dimensions $d=1$ and $2$, though in the marginal
case $d=2$ one has borderline phenomena such as weak localization.
Interest in the Anderson transition has been renewed by reports of 
metallic behavior in dilute two-dimensional electron-hole 
systems~\cite{rmp01}. While phenomenological, percolation-based
theories have been able to reproduce experimentally-observed trends in some
detail~\cite{meir}, attempts to reconcile basic theoretical 
assumptions to experimental evidence have only met limited success so 
far. For instance, numerical evidence
has been produced~\cite{interact01} against the idea that
electron-electron interactions (not included in scaling theory) might
play a role in driving the two-dimensional transition~\cite{rmp01}.

Even when one confines oneself to the original Anderson picture
of non-interacting electrons in three-dimensional lattices, where
the existence of a transition is not questioned, progress towards
extracting reliable numerical estimates of critical quantities
has been remarkably hard~\cite{ps81,mkk81,mkk83,bkmk85,bsk87,cbs90,mk94}.
Recently, systematic consideration of irrelevant variables and non-linear 
corrections to 
single-parameter scaling~\cite{cardy,so99,osk99,sok00,smo01,sldq01} has 
helped produce results with a fairly reasonable claim to consistently narrow
error bars.

It is thus of interest to reexamine the basic methods which have been
used in the past 20 years, in conjunction with SPST, 
to study the Anderson localization problem. A step in this 
direction has been given in Ref.~\onlinecite{kb02}, whose authors
obtained wave functions via exact diagonalization, for both 
one-dimensional and finite, $L \times L$, two-dimensional systems. By
averaging over randomness, they obtained  probability distributions of 
wavefunction amplitudes on sites at varying distances from an arbitrary
origin. Such distributions were compared to predictions from 
SPST; though agreement was good in 
$d=1$, the two-dimensional results were in contradiction to the idea of a 
single localization length depending only on disorder intensity: 
instead, a clear logarithmic increase with distance, at fixed disorder, 
was found from their fits for that quantity.

In Ref.~\onlinecite{kb02}, finite-size effects were avoided in $d=2$ by 
considering disorder strengths such that the corresponding
localization length, as predicted by single-parameter theory, is $\lambda 
\lesssim 10$ (see, e.g., Refs.~\onlinecite{ps81,mkk83}), and using 
suitably large 
systems with $L=300$. On the other hand, numerous studies of the Anderson
transition are set up on quasi-one dimensional geometries, for ease of
application of transfer-matrix (TM) or recursive Green's functions 
methods~\cite{ps81,mkk81,mkk83,bkmk85,bsk87,mk94,so99,osk99,sok00,sldq01}; 
extrapolation to bulk behavior ($d=2$ or $3$ as the case may be) 
is then performed with help of finite-size scaling theory~\cite{bar83}.

Here we consider TM methods, applied both to strictly one-dimensional 
systems and  to strips of a square lattice. 
Traditionally the TM approach has been used to calculate
Lyapunov exponents (directly related to the localization 
length of SPST)~\cite{ps81,ranmat}, and quantities obtainable
from such exponents, e.g., conductances~\cite{jlp86,pa86}. In 
Sec.~\ref{method} we recall how wavefunction amplitudes may be 
estimated in the TM context, and illustrate our approach  in the 
simple $d=1$ case by rederiving the corresponding distributions
found in Ref.~\onlinecite{kb02}. In Sec.~\ref{twod}, an analogous
treatment is developed for strips of a two-dimensional square lattice, 
and numerical results are displayed and discussed. Conclusions and final 
remarks are given in Sec.~\ref{conc}.     

\section{Method and one-dimensional illustration}
\label{method}
We consider the site-disordered Anderson model, for which the 
tight-binding Hamiltonian is written as
\begin{equation}   
{\cal H} = \sum_i \varepsilon_i|i\rangle\,\langle i| +
V \sum_{\langle i,j\rangle}|i\rangle\,\langle j| \ \ \ ,
\label{eq:ham}
\end{equation}
where the site self-energies $\varepsilon_i$ are independent,
identically distributed random variables obeying a specified
distribution,
$\langle i,j\rangle$ denotes nearest-neighbor sites on a regular
lattice, and the
energy scale is set by the hopping matrix element, $V \equiv 1$.
Disorder intensity is given by the
width $W$ of the self-energy probability distribution,
taken here as rectangular (same as in Ref.~\onlinecite{kb02}):\par
\begin{equation}
P (\varepsilon_i) = 
\begin{cases}
{{\rm constant} \quad -W/2 \leq \varepsilon_i \leq +W/2 }\cr
{0\qquad\qquad\quad {\rm otherwise\ .}}
\end{cases}
\label{eq:prob}
\end{equation}
In the TM approach~\cite{ps81}, one considers the Hamiltonian 
Eq.~(\ref{eq:ham}) on a quasi-one dimensional $L^{d-1} \times 
N$ system, $N \gg L$. Denoting by $k= 1, \dots, N$ the successive 
cross-sections, and $i = 1, \dots L^{d-1}$ the respective positions of 
sites within each cross-section, an electronic wave function at energy $E$ 
is given in terms of its local amplitudes,
$\{\,a_{ik}(E)\,\}$, and tight-binding orbitals $|ik\rangle$, as:
\begin{equation}
\Psi_E =\sum_{ik} a_{ik}(E)\,|ik\rangle\  .
\label{eq:psi(E)}
\end{equation}
With a corresponding change of notation, Eq.~(\ref{eq:ham}) reads:
\begin{equation}
{\cal H} = \sum_{ik} \varepsilon_{ik}\,|ik\rangle\,\langle ik| +
 \sum_{\langle ik,i^\prime k^\prime\rangle}|ik\rangle\,\langle i^\prime
k^\prime| \ \ ,
\label{eq:H(ik)}
\end{equation}
where $\langle ik,i^\prime k^\prime\rangle$ stands for nearest-neighbor
pairs. Applying Eq.~(\ref{eq:H(ik)}) to Eq.~(\ref{eq:psi(E)}) gives the
recursion relation
\begin{equation}
a_{i,(k+1)} = (E-\varepsilon_{ik})\,a_{ik}-
a_{i,(k-1)}-\sum_{i^\prime} a_{i^\prime,k}\ \ ,
\label{eq:recrel}
\end{equation} 
where ${i^\prime}$  denotes nearest neighbors of $i$ within the same 
cross-section (the $E$-dependence is omitted for clarity). In matricial 
form,
\begin{equation}
\begin{pmatrix}{\psi_{k+1}}\cr{\psi_k}\end{pmatrix}=
\begin{pmatrix}{P_k}&{-I}\cr{I}&{0}\end{pmatrix}
\begin{pmatrix}{\psi_k}\cr{\psi_{k-1}}\end{pmatrix}\  ,\ 
\psi_k \equiv 
\begin{pmatrix}{a_{1,k}}\cr {a_{2,k}}\cr {\cdots}\cr
{a_{L^{d-1},k}}\end{pmatrix}\ \,
\label{eq:matrix}
\end{equation} 
where, considering, e.g., periodic boundary conditions across the $d-1$ 
transverse directions~\cite{ps81},
\begin{equation}
 {P_k} =\begin{pmatrix}{E-\varepsilon_{1,k}}&{-1}&{\ \, 0}&{ \cdots}&{ 
-1}\cr
{-1 }&{ E-\varepsilon_{2,k}}&{ -1}&{\cdots }&{\ 0}\cr
{\cdots }&{ \cdots }&{ }&{ }&{ -1}\cr
{-1 }&{\ 0 }&{ \cdots  }&{-1 }&{\ E-\varepsilon_{L^{d-1},k}}\end{pmatrix}
\label{eq:pmatrix}\ .
\end{equation}
The $ (2L^{d-1} \times 2L^{d-1})$ matrix $T_k \equiv 
\begin{pmatrix}{P_k}&{-I}\cr{I}&{0}\end{pmatrix}$ is symplectic, that 
is, its eigenvalues occur in pairs $\{ \alpha_i, \alpha_i^{-1}\},\ 
i=1, \dots, L^{d-1}$.   
As explained at length in Refs.~\onlinecite{ps81,mkk83,mk94,ranmat},
the matrix product $M_N=\prod_{k=1}^N T_k$ gives rise to the
eigenvalues $\exp \gamma_1\ \dots\ \exp \gamma_{L^{d-1}}$, where
the $\gamma_i$ are the Lyapunov characteristic exponents (LCE)
for the problem,
setting the asymptotic divergence of the corresponding eigenvectors $v_i$.
Being a product of symplectic matrices, $M_N$ also has this property,
therefore the LCE occur in symmetric pairs $\{\gamma_i,-\gamma_i\}$.
Attention usually concentrates on the LCE of smallest modulus, 
$\gamma_{L^{d-1}}$, whose inverse
gives the longest decay length  (identified with the localization
length of the Anderson problem).

To see the meaning of the eigenvectors $v_i$, recall that
the physically acceptable (non-diverging) wave function is the one
associated with the {\em negative} LCE of smallest 
modulus, $\gamma_{L^{d-1}+1}$~\cite{souillard}; the corresponding 
tight-binding amplitudes 
$\{a_{ik}^{(L^{d-1}+1)}\}$ will give information on the shape of
the electronic wave function of interest (because of the symplectic 
character of the TM, one might equally consider the {\em inverse} of the 
amplitudes associated to  $\gamma_{L^{d-1}}$; however, in  
practice the amount of calculational effort is the same either way) . 
Little use appears to have been
made of this property in the context of TM studies of Anderson 
localization,
except for a calculation of lateral transport properties in layered 
media~\cite{chan}.

It is important to recall that  the site amplitudes are not the directly
relevant quantities in the TM approach; instead, in the quasi-one 
dimensional systems used here the
wavefunction decay rate must be defined by comparing the moduli of
suitable vectors, each with $2L^{d-1}$ 
components~\cite{ps81,mkk83,mk94}.

We now take strictly one-dimensional systems, and illustrate how the 
above ideas work. The matrices $T_k$ are $2 \times 2$,
and the relevant LCE is $\gamma_2$; the corresponding wavefunction 
amplitude at site $k$ is $a_k^{(2)}$. Starting with an arbitrary pair of 
states at neighboring sites, say $(a_0,a_1)=(1,1)$, we first iterate 
Eq.~(\ref{eq:matrix}) a number $N_{\rm in}$ of times, taking care to
orthonormalize the resulting vectors every $N_{\rm ortho}$ steps 
(typically, $N_{\rm in}=100$; we have used $N_{\rm ortho}=1$, but other 
authors have used $N_{\rm ortho} \simeq 10$ apparently without noticeable 
deterioration of results~\cite{mk94,ranmat}). With such initialization
the starting vectors are rotated in Hilbert space
towards the asymptotic direction of the eigenvectors of $M_N$. Having done
this, we rename the current site as the origin. Recalling from 
Eq.~(\ref{eq:matrix}) that the vectors being iterated involve both
$\psi_k$ and $\psi_{k+1}$, one sees that the appropriate quantities to
keep track of are the 
$b_k^{(2)} \equiv \{[a_k^{(2)}]^2+[a_{k+1}^{(2)}]^2\}^{1/2}$.
One might visualize the process as follows: starting from
the pair of site amplitudes $(a_0,a_1)$, one iteratively obtains the
pair $(a_1,a_2)$ and so
on, until (after $r$ iterations of $T$) one gets the pair
$(a_r,a_{r+1})$. This latter is legitimately said to be at a distance $r$
from the origin, that is, from  the original pair of sites.

\begin{figure}
{\centering \resizebox*{3.4in}{!}{\includegraphics*{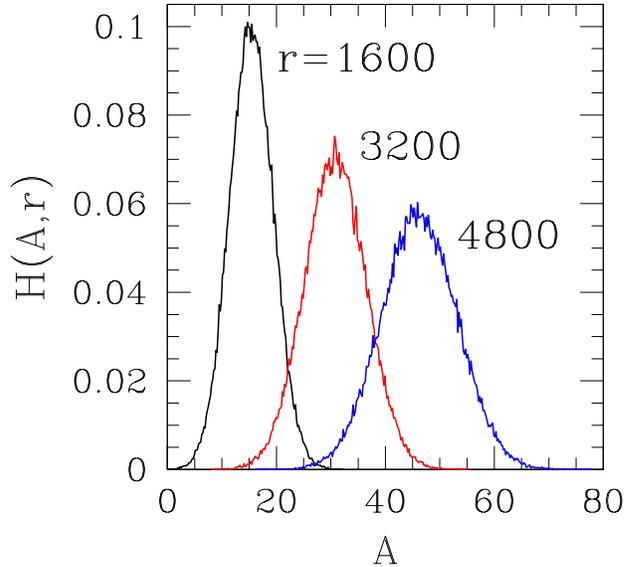}} \par}
\caption{Normalized histograms of occurrence of the logarithmic
decay factor, $A(r)$ of Eq.~(\protect{\ref{eq:A(r)}}), in $d=1$,
for $E=0$, $W=1.0$, and distances $r$ as shown. $N_{\rm s}=10^5$
samples were collected, for each $r$ (see text).}
\label{fig:fig1}
\end{figure}

We then start to accumulate 
the products of successive amplitudes $b_k^{(2)}$ at each $N_{\rm ortho}$ 
steps ( {\em after} orthogonalization, but {\em before} normalization).
At distance $r$ from 
the new origin, the (relative) wavefunction amplitude is given by
\begin{equation}
A(r) \equiv -\ln\frac{\psi^\prime (r)}{\psi^\prime (0)} =-\sum_{k=1}^{r} 
\ln b_k^{(2)}
\ , 
\label{eq:A(r)}
\end{equation}
where the notation of Ref.~\onlinecite{kb02} is used for ease of 
comparison, $\psi^\prime (k) \equiv \{[\psi_k]^2+[\psi_{k+1}]^2\}^{1/2}$,
and the fact that we always make $N_{\rm ortho}=1$
has been taken into account. In 
order to generate statistics of the $A(r)$ for a set of
distances $\{ r_1 < r_2 < \dots < r_{\rm max} \}$, one iterates the TM
for $N_0 \geq r_{\rm max}$ steps, collecting data at the specified
points; to collect the next sample, it suffices to keep iterating for 
another $N_0$ steps, with no need to reinitialize the wave functions, and 
so on. After a total of $N_{\rm in}+N_{\rm s}\,N_0$ iterations of the TM,
one has $N_{\rm s}$ samples of the $A(r)$ for each distance of interest,
The corresponding histograms $H(A,r)$ for 
$E=0$,  disorder strength $W=1.0$, $r=1600$, $3200$, and $4800$, and 
$N_{\rm s}=10^5$ are shown in 
Fig.~\ref{fig:fig1}. These values of the parameters were chosen in order
to enable comparison with the exact diagonalization results displayed in 
Figure 1 of Ref.~\onlinecite{kb02}.
Indeed, one finds excellent visual agreement between the respective data 
sets. Numerical analysis of the first three moments of our histograms
shows that: (i) for all three distances, they are well-fitted by gaussians 
of the form
\begin{equation}
H(A,r) = \left(\frac{\lambda}{\pi\sigma r}\right)^{1/2}\,\exp\left[\,
-\frac{\left(A-r/\lambda\right)^2}{\sigma r/\lambda}\,\right]\ ,
\label{eq:gfit1d}
\end{equation}  
with $\sigma \simeq 2.08$, $\lambda \simeq 104$, which are in
rather good agreement with the expected values from SPST~\cite{kb02,kw81}
 in the limit $N_{\rm s}, r \to \infty$, respectively $\sigma=2$, 
$\lambda =105.045/W^2$; and (ii) the dimensionless skewness $S$, defined 
as~\cite{numrec}:
\begin{equation}
 S \equiv \left\langle \left(\frac{ x - \langle x \rangle}{ 
\Delta}\right)^3 \right\rangle
\label{eq:skewdef}
\end{equation}
\noindent for a distribution with mean $ \langle x \rangle $ and
dispersion $ \Delta $, has the following values: $S=0.0250$, $0.0190$,
$0.0128$ respectively for $r=1600$, $3200$, $4800$.

Point (ii) is further indication that the
wavefunction amplitude distribution indeed approaches a log-normal 
shape (zero skewness), but only as $r \to \infty$. The approximate 
dependence on $r$ may be inferred as $S \sim r^{-1/2}$, from the 3 data
just quoted.

\section{Strips of a two-dimensional lattice}
\label{twod}
We now extend our approach to strips of a square lattice. For a strip
of width $L$, the matrices $T_k$ are $2 L \times 2 L$,
and the relevant LCE is $\gamma_{L+1}$; the corresponding wavefunction 
amplitudes at column $k$ are $a_{ik}^{(L+1)}$, $i=1, \dots, L$. 
Taking into account the normalization of Eq.~(\ref{eq:matrix}),
the appropriate decay factor here is 
\begin{equation}
A(r) =-\sum_{k=1}^{r} \ln 
\left(\,\sum_{j=k}^{k+1}\sum_{i=1}^L\left[\,a_{ij}^{(L+1)}
\,\right]^2\,\right)^{1/2}
\ \ (d=2).
\label{eq:A(r)2d}
\end{equation}
It must be stressed that Eq.~(\ref{eq:A(r)2d}) is not meant to imply 
an averaging process over site amplitudes $a_{ij}^{(L+1)}$; as remarked
above, these $2L$ quantities are not the directly relevant ones.
Instead, they give the modulus of the eigenvector associated to the
negative LCE of smallest absolute value~\cite{souillard}, whose decay
is to be followed.   

We have considered strips of even widths
$4 \leq L \leq 64$ sites and periodic boundary conditions across, and 
taken the disorder intensity $W=10$, in order to make contact with 
analogous results in  Ref.~\onlinecite{kb02}. For this value of $W$ 
SPST predicts the localization length to be $\lambda \simeq 
5.45$~\cite{mkk83}. SPST, together with finite-size scaling~\cite{bar83}
would imply that
\begin{equation}
H(A,r,L,W) = f\,\left(A,\frac{r}{L},\frac{r}{\lambda}\right)\ .
\label{eq:spst2d}
\end{equation}
In order to infer the $d=2$ behavior, one must consider the regime
$r$, $L \gg 1$, $r/L \lesssim 1$~\cite{bar83}. Although TM methods
make it easy to explore long distances ($r \gg L$) along the ``infinite''
direction, this fact is not directly relevant here, as the
corresponding regime would be one where strictly one-dimensional features
emerge.

In the following, we shall restrict ourselves to $r/L=1/2$ and $1$; 
according to Eq.~(\ref{eq:spst2d}), for 
each  value of $r/L$ one should then be able to collapse all
distributions against $r/\lambda$.

\begin{figure}
{\centering \resizebox*{3.4in}{!}{\includegraphics*{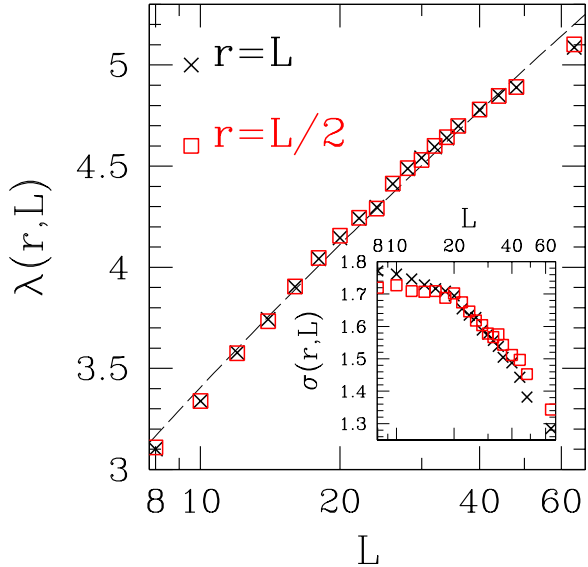}} \par}
\caption{Effective localization lengths on strips of $d=2$ lattice,
fitted from first and second moments of distributions to the gaussian
form,  Eq.~(\protect{\ref{eq:gfit1d}}),
for $E=0$, $W=10.0$.  Strip widths $L$ and distances $r$ as shown. 
$N_{\rm s}=5 \times 10^4$
samples for $L \leq 40$, $2\times 10^4$ otherwise. Dashed line is 
$\lambda=3.4\,(\log L)^{0.72}$ (see text). Scale is logarithmic on
horizontal axis. 
Inset: width of distribution, $\sigma$, as defined in
Eq.~(\protect{\ref{eq:gfit1d}}). Axis scales and symbols as in main
Figure. 
}
\label{fig:fig2}
\end{figure}
Again, we have examined the first three moments of the distributions
thus generated. Using only the first two, we have fitted data to
gaussians in the manner of Eq.~(\ref{eq:gfit1d}), for which the
effective  localization lengths $\lambda(r,L)$ are displayed in
Fig.~\ref{fig:fig2}. It can be seen that, for given $L$, the 
$\lambda$ are essentially the same both for $r=L/2$ and $L$. This shows 
that crude finite-size distortions do not play a role for the ranges
of $r$ and $L$ used. On the other hand, similarly to the findings of
Ref.~\onlinecite{kb02} and against the prediction of SPST, there is no 
single value of $\lambda$ to fit all distributions; instead, it 
grows with increasing $r$. However, our result differs from that of 
Ref.~\onlinecite{kb02}, in that the dependence of $\lambda$ is clearly  
not linear in $\log L$. This should not be seen as a direct contradiction, 
as the quantities under study are not identical (as was the case in $d=1$ 
): though they represent the same physical phenomenon of wavefunction 
decay, they do so in rather different geometries.

As regards the width of distribution $\sigma$ (again taking 
Eq.~(\ref{eq:gfit1d}) as a starting point) our results,
displayed in
the inset of Fig.~\ref{fig:fig2}, exhibit numerical values not unlike
those found in Ref.~\onlinecite{kb02}, in the sense of being consistently
smaller than the SPST prediction $\sigma=2$, but with the same order of
magnitude. One might expect that, for larger $r$, $L$ the decreasing
trend observed for $20 \lesssim L \lesssim 60$ would stabilize
close to $\sigma \simeq 1.3$ quoted in Ref.~\onlinecite{kb02}.
This, however, we have no means to ascertain at present.

We have not been able 
to fit the full range of data by a single power of either
$L$ or $\log L$; assuming, e.g., $\lambda \sim (\log L)^x$, the best
result from a nonlinear least-squares fit gives $x \simeq 0.72$,
corresponding to the
dashed line in Fig.~\ref{fig:fig2}. It is evident, from the Figure, that
the trend for large $L$ is towards an even slower variation.   

At this point, one might speculate that $\lambda$, as given by the 
gaussian fits, could eventually saturate for larger $L$, at a value
which might even be close to the SPST prediction. However, we shall now 
show that the gaussian fits themselves become increasingly unable to
reflect the properties of the wavefunction amplitude distributions.

\begin{figure}
{\centering \resizebox*{3.4in}{!}{\includegraphics*{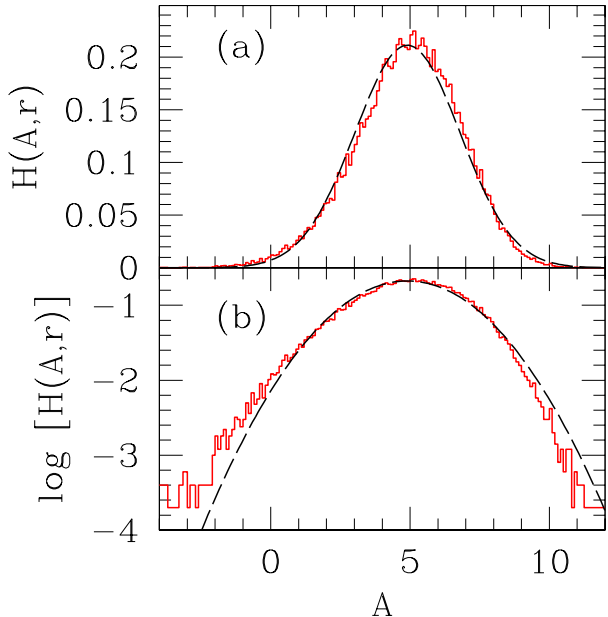}} \par}
\caption{Normalized histogram of occurrence of  $A(r)$ of 
Eq.~(\protect{\ref{eq:A(r)2d}}) for $E=0$, $W=10.0$, $L=48$, $r=24$
(full line). Skewness=$-0.288$. Dashed line: Gaussian fit 
(Eq.~(\protect{\ref{eq:gfit1d}})), using first and second moments of
distribution. Vertical scale is linear in (a), and logarithmic in (b),
the latter in order to emphasize discrepancies between data and fit at
the extremes.}
\label{fig:fig3}
\end{figure}

Indeed, we have found that the skewness of distributions is negative,
and {\em increases} in absolute value as $r$, $L$ grow.
As an example, Fig.~\ref{fig:fig3} shows the raw data for $L=48$, 
$r=24$, together with the
corresponding  gaussian fit of
Eq.~(\ref{eq:gfit1d}), obtained using the first and second moments 
of the distribution. The effect of negative skewness is apparent
in that the gaussian approximation overshoots the data for large $A$
(i.e., predicts a {\em small} amplitude to occur {\em more frequently} 
than
observed in fact), and  undershoots for small $A$ (predicts a {\em large} 
amplitude to occur {\em less frequently} than observed). 

\begin{figure}
{\centering \resizebox*{3.4in}{!}{\includegraphics*{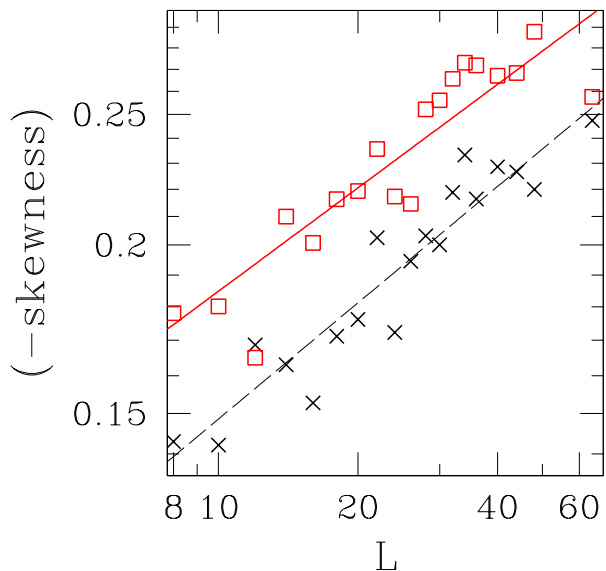}} \par}
\caption{Double-logarithmic plots of negative skewness of wavefunction
amplitude distributions, $A(r)$ of 
Eq.~(\protect{\ref{eq:A(r)2d}}), for $E=0$, $W=10.0$ against strip width
$L$, and corresponding least-squares fits to single-power forms.
Crosses and dashed line: $r=L$. Squares and full line: $r=L/2$.}
\label{fig:fig4}
\end{figure}

Skewness data for the ranges of $r$ and $L$ used here are displayed
in Fig.~\ref{fig:fig4}, together with fits of single-power forms, 
$-S \sim L^x$, for the subsets corresponding respectively to $r=L$
(dashed line, $x \simeq 0.29$) and $r=L/2$
(full line, $x \simeq 0.25$). Despite the large amount of scatter,
the increasing trend against growing $r$, $L$ is unmistakably present.
This means, in turn, that for larger and larger systems the
distributions become ever less amenable to fitting by gaussians,
as predicted by SPST. This would remain true even if a hypothetical
saturation should occur for values of $r$ and $L$ larger than those 
investigated here. 

Recall, from Eq.~(\ref{eq:skewdef}), that skewness is 
invariant under the usual histogram-collapse rescaling~\cite{kb02},
$H_s(A_s,r)=H(A,r)\,\sqrt{\pi\sigma r/\lambda (r)}$, where the shifted
variable is $A_s \equiv \left[ A-r/\lambda (r)\right]/\sqrt{\sigma 
r/\lambda (r)}$. Therefore, this is a legitimate extra parameter
to characterize the distributions. Similar results were found 
experimentally, for the conductance distribution in quasi-one
dimensional gold wires~\cite{mw02}.

Negative skewness of wavefunction amplitude distributions works 
in the same way as (in the limited context of gaussian fits) does the    
finding that $\lambda(r)$ increases with $r$: both contribute to a 
slower decay of electronic wave functions, compared with
the constant-$\lambda$, zero-skewness, SPST picture. Of course, the 
evidence just presented is not enough to argue that there must be a
localization-delocalization transition in $d=2$; the idea that this
is the borderline dimensionality, as predicted by SPST, most likely holds 
true. Nonetheless, we have shown robust evidence for deviations from SPST 
in $d=2$, whose consequences still have to be worked out in full,  

\section{Conclusions}
\label{conc}
We have made use of suitable properties of the vectors which are iterated 
in the TM approach to Anderson localization~\cite{ps81,mkk83,mk94,ranmat}, 
in order to generate the statistical distribution of electronic 
wavefunction amplitudes at  sites of $L^{d-1} \times \infty$ 
disordered systems. We have considered $d=1$ (for which our approach 
is shown to reproduce the exact diagonalization results of 
Ref.~\onlinecite{kb02}), and $d=2$. In the latter case, since
the $L \times \infty$ geometry of our systems
differs from that ($L \times L$) of  Ref.~\onlinecite{kb02},
a perfect match is not to be expected; however, some basic physical 
properties are found to hold for both cases. In particular, attempted
fits of gaussian (log-normal) forms to the wavefunction amplitude 
distributions result 
in effective localization lengths growing with distance, contrary to
the SPST prediction. We have gone further, and shown that the 
distributions possess a negative skewness, which  
is invariant under the usual histogram-collapse rescaling, and
increases with distance (at least for the range of parameters used
in our study). 

Such deviations from the expected behavior are evidence of 
slower decay 
of electronic wavefunctions than predicted by SPST; it still must be
worked out whether or not some phenomena specific to $d=2$, such as
weak localization, or the recently-observed metal-insulator transition in
dilute two-dimensional dilute electron-hole systems~\cite{rmp01},
carry the fingerprints of the anomalies reported here. We expect that
the present results may motivate further work along these lines.

\begin{acknowledgments}

Research of S.L.A.d.Q. is partially supported by the Brazilian agencies
CNPq (Grant No. 30.1692/81.5), FAPERJ (Grants
Nos. E26--171.447/97 and E26--151.869/2000) and FUJB-UFRJ.
\end{acknowledgments}

\end{document}